\documentclass[aoas,nameyear,seceqn,dvips]{arximspdf}
\usepackage{stfloats}
\usepackage{graphicx}

\doi{10.1214/09-AOAS321}
\volume{4}
\issue{2}
\pubyear{2010}
\firstpage{615}
\lastpage{644}

\makeatletter
\newproclaim{barf}{Bayesian analogical reasoning formulation}
\fnbelowfloat
\makeatother

\begin{document}
\begin{frontmatter}

\title{Ranking relations using analogies in biological and information networks\thanksref{M1}}
\runtitle{Ranking relations using analogies}
\thankstext{M1}{Supported in part by NSF Grant DMS-09-07009, by NIH Grant R01 GM096193,
and by the Gatsby Charitable Foundation.}

\begin{aug}
\author[a]{\fnms{Ricardo} \snm{Silva}\corref{}\ead[label=e1]{ricardo@stats.ucl.ac.uk}},
\author[b]{\fnms{Katherine} \snm{Heller}\ead[label=e2]{heller@gatsby.ucl.ac.uk}},\\
\author[b]{\fnms{Zoubin} \snm{Ghahramani}\ead[label=e3]{zoubin@eng.cam.ac.uk}}
\and
\author[c]{\fnms{Edoardo M.} \snm{Airoldi}\ead[label=e4]{airoldi@fas.harvard.edu}}
\runauthor{Silva, Heller, Ghahramani and Airoldi}
\affiliation{University College London, University of Cambridge,\\
University of Cambridge
and Harvard University}
\address[a]{
R. Silva\\
University College London\\
Gower Street\\
London, WC1E 6BT\\
United Kingdom\\
\printead{e1}}

\address[b]{
K. Heller\\
Z. Ghahramani\\
University of Cambridge\\
Trumpington Street\\
Cambridge, CB2 1PZ\\
United Kingdom\\
\printead{e2}\\
\phantom{E-mail:\ }\printead*{e3}}

\address[c]{
E. M. Airoldi\\
Harvard University\\
1 Oxford street\\
Cambridge, Massachusetts 02138\\
USA\\
\printead{e4}}
\end{aug}

\received{\smonth{5} \syear{2009}}
\revised{\smonth{11} \syear{2009}}

%
\begin{abstract}
Analogical reasoning depends fundamentally on the ability to learn and
generalize about relations between objects.\vspace*{1pt} We develop an approach to
relational learning which, given a set of pairs of objects $\mathbf{S}
= \{A^{(1)}\dvtx B^{(1)}, A^{(2)}\dvtx B^{(2)}, \ldots,
A^{(N)}\dvtx B^{(N)}\}$,
measures how well other pairs $A\dvtx B$ fit in with the set~$\mathbf{S}$.
Our work addresses the following question: is the relation between
objects $A$ and $B$ analogous to those relations found in
$\mathbf{S}$? Such questions are particularly relevant in information
retrieval, where an investigator might want to search for analogous
pairs of objects that match the query set of interest. There are many
ways in which objects can be related, making the task of measuring
analogies very challenging. Our approach combines a similarity measure
on function spaces with Bayesian analysis to produce a ranking. It
requires data containing features of the objects of interest and a
link matrix specifying which relationships exist; no further
attributes of such relationships are necessary. We illustrate the
potential of our method on text analysis and information networks. An
application on discovering functional interactions between pairs of
proteins is discussed in detail, where we show that our approach can
work in practice even if a small set of protein pairs is provided.
\end{abstract}
%
%
\begin{keyword}
\kwd{Network analysis}
\kwd{Bayesian inference}
\kwd{variational approximation}
\kwd{ranking}
\kwd{information retrieval}
\kwd{data integration}
\kwd{\textit{Saccharomyces cerevisiae}}.
\end{keyword}

\end{frontmatter}
%
\section{Contribution}

Many university admission exams, such as the American Scholastic
Assessment Test (SAT) and Graduate Record Exam (GRE), have
historically included a section on analogical reasoning. A
prototypical analogical reasoning question is as follows:

\begin{enumerate}[(A)]
\item[] $\mathtt{doctor \dvtx hospital}$:
\item[\texttt{(A)}] \texttt{sports} $\mathtt{fan \dvtx stadium}$\vadjust{\goodbreak}
\item[\texttt{(B)}] \texttt{cow} $\mathtt{\dvtx farm}$
\item[\texttt{(C)}] $\mathtt{professor \dvtx college}$
\item[\texttt{(D)}] $\mathtt{criminal \dvtx jail}$
\item[\texttt{(E)}] $\mathtt{food \dvtx grocery}$ \texttt{store}
\end{enumerate}

The examinee has to answer which of the five pairs best matches the
relation implicit in $\mathtt{doctor\dvtx hospital}$. Although all candidate
pairs have some type of relation, pair $\mathtt{professor\dvtx college}$ seems
to best fit the notion of (\textit{profession}, \textit{place of
work}), or
the ``works in'' relation implicit between doctor and hospital.

This problem is nontrivial because measuring the similarity between
objects directly is not an appropriate way of discovering
analogies, as extensively discussed in the cognitive
science literature. For instance, the analogy between an electron spinning
around the nucleus of an atom and a planet orbiting around the Sun is not
justified by isolated, nonrelational, comparisons of an electron to a
planet, and of an atomic nucleus to the Sun [\citet{gentner83}].
Discovering the underlying relationship between the
elements of each pair is key in determining analogies.

\subsection{Applications}

This paper concerns practical problems of data analysis where
analogies, implicitly or not, play a role. One of our motivations
comes from the \textit{bioPIXIE}\setcounter{footnote}{1}\footnote{\url{http://pixie.princeton.edu/pixie/}.} project
[\citet{myers05}]. bioPIXIE is a tool for exploratory analysis
of protein--protein interactions. Proteins have \textit{multiple
functional roles} in the cell, for example, regulating metabolism and regulating
cell cycle, among others. A protein often assumes different functional
roles while interacting with different proteins. When a molecular
biologist experimentally observes an interaction between two proteins,
for example, a binding event of $\{P_i, P_j\}$, it might not be clear which
function that particular interaction is contributing to.
The bioPIXIE system allows a molecular biologist to input a set
$\mathbf{S}$ of proteins that are believed to have a particular
functional role in common, and generates a list of other proteins that
are deduced to play the same role. Evidence for such predictions is
provided by a variety of sources, such as the expression levels for
the genes that encode the proteins of interest and their cellular
localization. Another important source of information bioPIXIE takes
advantage of is a matrix of relationships, indicating which proteins
interact according to some biological criterion. However, we do not
necessarily know which interactions correspond to which functional
roles.

The application to protein interaction networks that we develop in
Section~\ref{secexp-ppi} shares some of the features and motivations
of bioPIXIE. However, we aim at providing more detailed
information. Our input set $\mathbf{S}$ is a \textit{small set of pairs}
of proteins that are postulated to all play a common role, and we want
to rank \textit{other pairs} $P_i\dvtx P_j$ according to how similar they
are with respect to $\mathbf{S}$. The goal is to automatically return
pairs that correspond to analogous interactions.

To use an analogy itself to explain our procedure, recall the SAT
example that opened this section. The pair of words
${\mathtt{doctor}\dvtx \mathtt{hospital}}$ presented in the SAT question play the role of a
protein--protein interaction and is the smallest possible case of
$\mathbf{S}$, that is, a single pair. The five choices \texttt{A}--\texttt{E}
in the SAT
question correspond to other observed protein--protein interactions we
want to match with~$\mathbf{S}$, that is, other possible pairs. Since
multiple valid answers are possible, we rank them according to a
similarity metric. In the application to protein interactions, in
Section~\ref{secexp-ppi}, we perform thousands of queries and we
evaluate the goodness of the resulting rankings according to multiple
gold standards, widely accepted by molecular and cellular biologists
[\citet{Ashbetal2000}; \citet{kanegoto2000}; \citet{mewes04}].

The general problem of interest in this paper is a practical problem of
\textit{information retrieval}
[\citet{manning08}] for exploratory data analysis:
given a \textit{query set} $\mathbf{S}$ of linked pairs, which other pairs
of objects in my relational database are linked in a similar way? We
apply this analysis to cases where it is not known how to explicitly
describe the different classes of relations, but good models to
predict the \textit{existence} of relationships are available.
In Section~\ref{secexp-webkb} we consider an application to
information retrieval in text documents for illustrative
purposes. Given a set of pairs of web pages which are related by
some hyperlink, we would like to find other pairs of pages that are
linked in a similar way.
In information network settings, the proposed method could be useful,
for instance, to answer queries for encyclopedia pages relating
scientists and their major discoveries, to search for analogous
concepts, or to identify the absence of analogous concepts, in
Wikipedia. From an evaluation perspective, this application domain
provides an example where large scale evaluation is more
straightforward than in the biological setting.

In this paper we introduce a method for ranking relations
based on the Bayesian similarity criterion underlying
\textit{Bayesian sets}, a method originally proposed by \citet{gha05} and
reviewed in Section~\ref{secreview}. In contrast to Bayesian sets,
however, our method is tailored to drawing analogies between pairs of
objects. We also provide supplementary material with a Java implementation
of our method, and instructions on how to rebuild the experiments
[\citet{silvasup10}].

\subsection{Related work}

To give an idea of the type of data which our method is useful for
analyzing, consider the methods of \citet{turney05} for automatically
solving SAT problems. Their analysis is based on a large corpus of
documents extracted from\vadjust{\goodbreak} the World Wide Web. Relations between two
words~$W_i$ and~$W_j$ are characterized by their joint co-ocurrence
with other relevant words (such as particular prepositions) within a
small window of text. This defines a set of features for each
$W_i\dvtx W_j$ relationship, which can then be compared to other pairs of
words using some notion of similarity. Unlike in this application,
however, there are often no (or very few) explicit features for the
relationships of interest. Instead we need a method for defining
similarities using features of the objects in each relationship, while
at the same time avoiding the mistake of directly comparing objects
instead of relations.

One of the earliest approaches for
determining analogical similarity was introduced by
\citet{rumelhart73}. In their paper, one is initially given a set of pairwise
distances between objects (say, by the subjective judgement of a group
of people). Such distances are used to embed the given objects in a
latent space via a multidimensional scaling approach. A related pair
$A\dvtx B$ is then represented as a vector connecting $A$ and $B$ in the
latent space. Its similarity with respect to another pair $C\dvtx D$ is
defined by comparing the direction and magnitude of the corresponding
vectors. Our approach is probabilistic instead of geometrical, and
operates directly on the object features instead of pairwise
distances.

We will focus solely on ranking pairwise relations. The idea can be
extended to more complex relations, but we will not pursue this here.
Our approach is described in detail in Section~\ref{secmodel}.

Finally, the probabilistic, geometrical and logical approaches applied
to analogical reasoning problems can be seen as a type of
relational data analysis [\citet{dzeroski01}; \citet{getoor07}]. In
particular, analogical reasoning is a part of the more general problem
of generating latent relationships from relational data. Several
approaches for this problem are discussed in Section
\ref{secpreviouswork}. To the best of our knowledge, however, most
analogical reasoning applications are interesting proofs of concept
that tackle ambitious problems such as planning
[\citet{veloso93}], or are motivated as models of cognition
[\citet{gentner83}]. Our goal is to create an off-the-shelf method for
practical exploratory data analysis.

\section{A review of probabilistic information retrieval and the
Bayesian sets method}
\label{secreview}

The goal of information retrieval is to provide data points (e.g.,
text documents, images, medical records) that are judged to be
relevant to a particular query. Queries can be defined in a variety
of ways and, in general, they do not specify exactly which records
should be presented. In practice, retrieval methods rank data points
according to some measure of similarity with respect to the query
[\citet{manning08}]. Although queries can, in practice, consist of any
piece of information, for the purposes of this paper we will assume
that queries are sets of objects of the same type we want to retrieve.

Probabilities can be exploited as a measure of similarity. We will
briefly review one standard probabilistic framework for information
retrieval [\citet{manning08}, Chapter 11]. Let $R$ be a binary random
variable representing whether an arbitrary data point $X$ is
``relevant'' for a given query set $\mathbf{S}$ ($R = 1$) or not ($R
= 0$). Let $P(\cdot\vert \cdot)$ be a generic probability mass function or
density function, with its meaning given by the context.
Points are ranked in decreasing order by the following
criterion:
\[
\frac{P(R = 1\vert X, \mathbf{S})}{P(R = 0\vert X, \mathbf{S})}
=
\frac{P(R = 1\vert \mathbf{S})}{P(R = 0\vert \mathbf{S})}
\frac{P(X \vert R = 1, \mathbf{S})}{P(X \vert R = 0, \mathbf{S})},
\]
which is equivalent to ranking points by the expression
%
%
\begin{equation}
\label{eqgeneral-score}
\log P(X \vert R = 1,\mathbf{S}) - \log P(X \vert R = 0, \mathbf{S}).
\end{equation}

The challenge is to define what form $P(X \vert R = r, \mathbf{S})$ should
assume. It is not practical to collect labeled data in advance which,
for every possible class of queries, will give an
estimate for $P(R = 1\vert X, \mathbf{S})$: in general, one cannot
anticipate which classes of queries will exist. Instead, a variety of approaches
have been developed in the literature in order to define a suitable
instantiation of (\ref{eqgeneral-score}). These include a method that builds
a classifier on-the-fly using $\mathbf{S}$ as elements of the positive class
$R = 1$, and a random subset of data points as the negative class $R = 0$
[e.g., \citet{turney08u}].
%

\begin{figure}

\includegraphics{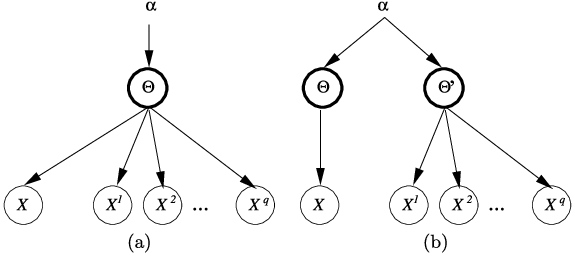}

\caption{In order to score how well an arbitrary element $X$ fits in
with query set $\mathbf{S} = \{X^1, X^2, \ldots, X^q\}$, the Bayesian sets
methodology compares the marginal likelihood of the model in \textup{(a)},
$P(X, \mathbf{S})$, against the model in \textup{(b)}, $P(X)P(\mathbf{S})$. In
\textup{(a)}, the random parameter vector~$\Theta$ is given a prior defined by the
(fixed) hyperparameter $\alpha$. The same (latent) parameter vector is shared
by the query set and the new point. In \textup{(b)}, the parameter vector
$\Theta$ that generates $X$ is different from the one that generates
the query set.}
\label{figbsets}
\end{figure}

The Bayesian sets method of \citet{gha05} is a state-of-the-art
probabilistic method for ranking objects, partially inspired by
Bayesian psychological models of generalization in human cognition
[\citet{tenenbaum01}]. In this setup the event ``$R = 1$'' is equated
with the event that $X$ and the elements of $\mathbf{S}$ are i.i.d.
points generated by the same model. The event ``$R = 0$'' is the event
by which $X$ and $\mathbf{S}$ are generated by two independent models:
one for~$X$ and another for $\mathbf{S}$. The parameters of all models
are random variables that have been integrated out, with fixed (and
common) hyperparameters. The result is the instantiation of
(\ref{eqgeneral-score}) as
%
%
\begin{equation}
\label{eqbsets-score}
\log P(X \vert \mathbf{S}) - \log P(X) =
\log\frac{P(X, \mathbf{S})}{P(X)P(\mathbf{S})},
\end{equation}
the Bayesian sets \textit{score function} by which we rank points $X$
given a
query $\mathbf{S}$. The right-hand side was rearranged to provide a more
intuitive
graphical model, shown in Figure~\ref{figbsets}. From this graphical
model interpretation we can see that the score function
is a Bayes factor comparing two models [\citet{KassRaft1995}].

In the next section we describe how the Bayesian sets method can be
adapted to define analogical similarity in the biological and
information networks settings we consider, and why such modifications
are necessary.

\section{A model of Bayesian analogical similarity for relations}
\label{secmodel}

To define an analogy is to define a measure of similarity between
structures of related objects. In our setting, we need to measure the
similarity between\vadjust{\goodbreak} pairs of objects. The key aspect that
distinguishes our approach from others is that we focus on the
similarity between \textit{functions} that map pairs to links, rather
than focusing on the similarity between the \textit{features} of objects
in a candidate pair and the features of objects in the query pairs.

As an illustration, consider an analogical reasoning question from a
SAT-like exam where for a given pair (say, $\mathit{water\dvtx river}$) we have
to choose, out of 5 pairs, the one that best matches the type of
relation implicit in such a ``query.'' In this case, it is reasonable
to say $\mathit{car\dvtx highway}$ would be a better match than (the somewhat
nonsensical) $\mathit{soda\dvtx ocean}$, since cars flow on a highway, and so
does water in a~river. Notice that if we were to measure the
similarity between \textit{objects} instead of \textit{relations},
$\mathit{soda\dvtx ocean}$ would be a much closer pair, since \textit{soda} is similar
to \textit{water}, and \textit{ocean} is similar to \textit{river}.

Nevertheless, it is legitimate to infer relational similarity from
individual object features, as summarized by \citet{gentner98} in
their ``kind world hypothesis.'' What is needed is a mechanism by which
object features should be weighted in a particular relational
similarity problem. We postulate that, in analogical reasoning,
similarity between features of objects is only meaningful to the
extent by which such features are useful to predict the existence of
the relationships.

Our approach can be described as follows. Let $\mathcal A$ and
$\mathcal B$ represent
object spaces.\vspace*{1pt} To say that an interaction $A\dvtx B$ is analogous to
$\mathbf{S} = \{A^{(1)}\dvtx B^{(1)},\break A^{(2)}\dvtx B^{(2)},
\ldots,
A^{(N)}\dvtx B^{(N)}\}$ amounts to implicitly defining a measure of
similarity\vspace*{1pt} between the pair $A\dvtx B$ and the set of pairs $\mathbf{S}$,
where each query item $A^{(k)}\dvtx B^{(k)}$ corresponds to some pair
$A^i\dvtx B^j$. However, this similarity is not directly derived from the
similarity of the information contained in the distribution of objects
themselves, $\{A^i\} \subset\mathcal A$, $\{B^i\} \subset\mathcal
B$. Rather, the similarity between $A\dvtx B$ and the set $\mathbf{S}$ is
defined in terms of the similarity of the \textit{functions} mapping the
pairs as being linked. Each possible function captures a different
possible relationship between the objects in the pair.

\begin{barf*} Consider a space of
latent functions in $\mathcal A \times\mathcal B \rightarrow\{0,
1\}$. Assume that $A$ and $B$ are two objects classified as linked by
some unknown function $f(A, B)$, that is, $f(A, B) = 1$. We want to
quantify how similar the function $f(A, B)$ is to the function
$g(\cdot, \cdot)$, which classifies all pairs $(A^i, B^j) \in\mathbf{S}$ as being linked, that is, where $g(A^i, B^j) = 1$. The similarity
should depend on the observations $\{\mathbf{S}, A, B\}$ and
our prior distribution over $f(\cdot,\cdot)$ and $g(\cdot,\cdot)$.
\end{barf*}

Functions $f(\cdot)$ and $g(\cdot)$ are unobserved, hence the need for
a prior that will be used to integrate over the function space. Our
similarity metric will be defined using Bayes factors, as explained
next.

\subsection{Analogy in function spaces via logistic regression}

For simplicity, we will consider a family of latent
functions that is parameterized by a finite-dimensional vector: the
logistic regression function with multivariate Gaussian priors for its
parameters.

For a particular pair $(A^i \in\mathcal A$, $B^j \in\mathcal B)$,
let $X^{ij} = [\Phi_1(A^i, B^j) \hspace*{12pt}\Phi_2(A^i, B^j)\break \cdots\hspace*{12pt} \Phi_K(A^i, B^j)]^\mathsf{T}$ be a
point on a feature space defined by the mapping $\Phi\dvtx \mathcal A
\times\mathcal B \rightarrow\Re^K$.
This feature space mapping computes a $K$-dimensional vector of
attributes of the pair that may be potentially relevant to predicting
the relation between the objects in the pair.
Let $L^{ij} \in\{0, 1\}$ be
an indicator of the existence of a link or relation between $A^i$ and
$B^j$ in
the database. Let $\Theta= [\theta_1, \ldots, \theta_{K}]^\mathsf{T}$
be the parameter vector for our logistic regression model such that
%
%
\begin{equation}
P(L^{ij} = 1 | X^{ij}, \Theta) = \operatorname{logistic}(\Theta^{\mathsf T} X^{ij}),
\end{equation}
where $\operatorname{logistic}(x) = (1 + e^{-x})^{-1}$.

We now apply the same score function underlying the Bayesian sets
methodology explained in Section~\ref{secreview}. However, instead of
comparing objects by marginalizing over the parameters of their
feature distributions, we compare \textit{functions} for link indicators
by marginalizing over the parameters of the functions.

Let $\mathbf{L^S}$ be the vector of link indicators for $\mathbf{S}$:
in fact, each $L \in\mathbf{L^S}$ has the value $L = 1$, indicating
that every pair of objects in
$\mathbf{S}$ is linked. Consider the following Bayes factor:
%
%
\begin{equation}
\label{eqrbsets-bayesfactor}
\frac{P(L^{ij} = 1, \mathbf{L^S} = 1\vert X^{ij}, \mathbf{S})}{P(L^{ij}
= 1\vert X^{ij})P(\mathbf{L^S} = 1\vert  \mathbf{S})}.
\end{equation}

\begin{figure}

\includegraphics{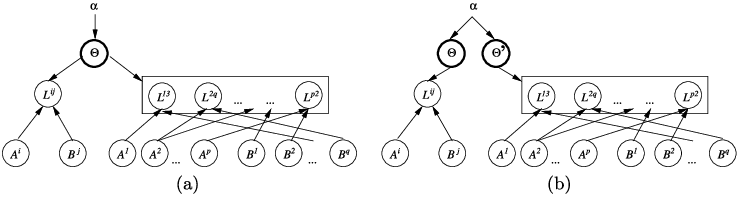}

\caption{The score of a new data point $\{A^i, B^j\}$ is given by the
Bayes factor that
compares models~\textup{(a)} and~\textup{(b)}. Node $\alpha$ represents the
hyperparameters for $\Theta$. In \textup{(a)}, the generative model is the same
for both the new point and the query set represented in the
rectangle. Notice that our conditioning set $\mathbf{S}$ of pairs
might contain repeated instances of a same point, that is, some $A$ or
$B$ might appear multiple times in different relations, as illustrated
by nodes with multiple outgoing edges. In \textup{(b)}, the new point and
the query set do not share the same parameters.}
\label{figbayes-factor}
\end{figure}

This is an adaptation of equation (\ref{eqbsets-score}) where
relevance is
defined now by whether $L^{ij}$ and $\mathbf{L^S}$ were generated by
the same model, for fixed $\{X^{ij}, \mathbf{S}\}$. In one sense, this
is a discriminative Bayesian sets model, where we predict links
instead of modeling joint object features. Since we are integrating
out $\Theta$, a prior for this parameter vector is needed. The
graphical models corresponding to this Bayes factor are illustrated in
Figure
\ref{figbayes-factor}.

Thus, each pair $(A^i, B^j)$ is evaluated with respect to a query
set $\mathbf{S}$ by the score function given in
(\ref{eqrbsets-bayesfactor}), rewritten after taking a logarithm and
dropping constants as
%
%
\begin{eqnarray}
\label{eqscore}
\operatorname{score}(A^i, B^j) &=& \log P(L^{ij} = 1\vert X^{ij}, \mathbf{S}, \mathbf
{L^S} = 1)
\nonumber
\\[-8pt]
\\[-8pt]
\nonumber
&&{}- \log P(L^{ij} = 1\vert X^{ij}).
\end{eqnarray}

The exact details of our procedure are as follows.
We are given a relational database $(\mathcal{D}_A, \mathcal{D}_B,
\mathcal{L}_{AB}$). Dataset $\mathcal D_A$ ($\mathcal D_B$) is a sample
of objects of type $\mathcal A$
($\mathcal B$). Relationship table $\mathcal{L}_{AB}$ is a binary
matrix modeled as generated from a logistic regression model of link
existence. A query proceeds according to the following steps:

\begin{enumerate}
\item the user selects a set of pairs $\mathbf{S}$ that are linked in
the database, where the pairs in $\mathbf{S}$ are assumed to have
some relation of interest;
\item the system performs Bayesian inference to obtain the
corresponding posterior
distribution for $\Theta$, $P(\Theta| \mathbf{S}, \mathbf{L^S})$,
given a Gaussian prior $P(\Theta)$;
\item the system iterates through all linked pairs, computing the
following for each pair:
\[
P(L^{ij} = 1 | X^{ij}, \mathbf{S}, \mathbf{L^S} = 1)=\int P(L^{ij} = 1 | X^{ij}, \Theta)P(\Theta| \mathbf{S}, \mathbf{L^S}
= 1 )\, d\Theta.
\]
$P(L^{ij} = 1 | X^{ij})$ is similarly computed by integrating over
$P(\Theta)$.
All pairs are presented in decreasing order according to the score in
equation (\ref{eqscore}).
\end{enumerate}

\begin{figure}

\includegraphics{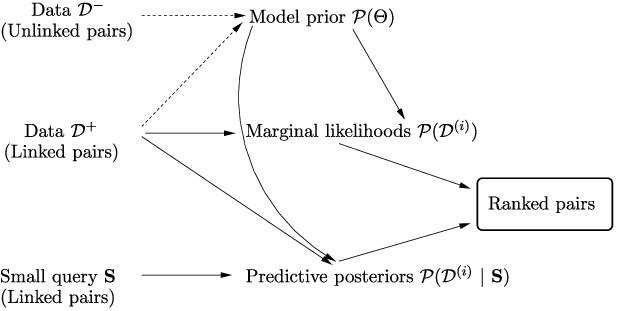}

\caption{General framework of the procedure: first, a ``prior'' over
parameters $\Theta$
for a link classifier is defined empirically using linked and unlinked
pairs of points (the dashed edges indicate that creating a prior
empirically is optional,
but in practice we rely on this method). Given a query set $\mathbf{S}$
of linked pairs of
interest, the system computes the predictive likelihood of each linked
pair $\mathcal D^{(i)} \in\mathcal D^+$ and compares it to the
conditional predictive likelihood, given the query. This defines a
measure of similarity
with respect to $\mathbf{S}$ by which all pairs in $\mathcal D^+$ are sorted.}
\label{figoverview}
\end{figure}

The integral presented above does not have a
closed formula. Because computing the integrals by a Monte Carlo
method for a large number of pairs would be unreasonable, we use a
variational approximation [\citet{JordGhahJaakSaul1999}; \citet{Airo2007b}].
Figure~\ref{figoverview} presents a summary of the approach.

%

The suggested setup scales as $O(K^3)$ with the feature space
dimension, due to the matrix inversions
necessary for (variational) Bayesian logistic regression
[\citet{jak00}]. A less precise approximation to $P(\Theta|
\mathbf{S}, \mathbf{L^S})$ can be imposed if the dimensionality of
$\Theta$ is too high. However, it is important to point out that once
the initial integral $P(\Theta\vert \mathbf{S}, \mathbf L^{\mathbf{S}})$
is approximated,
each score function can be computed at a cost of $O(K^2)$.

Our analogical reasoning formulation is a relational model in that it
models the presence and absence of interactions between objects. By
conditioning on the link indicators, the similarity score between
$A\dvtx B$ and $C\dvtx D$ is always a function of pairs $(A, B)$ and $(C,
D)$ that is not in general decomposable as similarities between~$A$
and $C$, and $B$ and $D$.

\subsection{Comparison with Bayesian sets and stochastic block models}
\label{secblock-comp}

The model presented in Figure~\ref{figbayes-factor} is a \textit{
conditional} independence model for relationship indicators, that is,
given object features and parameters, the entries of $\mathcal L_D$
are independent. However, the entries in $\mathcal L_D$ are in general
\textit{marginally} dependent. Since this is a model of relationships
given object attributes, we call the model introduced here
the \textit{relational Bayesian sets model}.

Our approach has some similarity to the so-called \textit{stochastic
block models}. These models were developed four decades ago in the
network literature to quantify the notion of ``structural
equivalence'' by means of blocks nodes that instantiate similar
connectivity patterns [\citet{LorrWhit1971}; \citet{HollLein1975}]. Modern
stochastic block model approaches, in statistics and machine learning,
build on these seminal works by introducing the discovery of the block
structure as part of the model search strategy
[\citet{FienMeyeWass1985}; \citet{NowiSnij2001};
\citet{kemp06}; \citet{XuTresYuKrie2006};
Airoldi et al. (\citeyear{AiroBleiXingFien2005,AiroBleiFienXing2008});
\citet{hoff2008}].
The observed features in our approach, $X^{ij}$, effectively play the
same role as the latent indicators in stochastic block
models.\footnote{In a stochastic block model, typically each object has
a single feature $\eta$ indicating membership to some latent class.
For a pair $A^i, B^j$, the corresponding feature vector $X^{ij}$ would
be $(\eta_A,
\eta_B)$.} Since $X^{ij}$
is observed, there is no need to integrate over the feature space to
obtain the posterior distribution of $\Theta$. This computational
efficiency is particularly
relevant in information retrieval and exploratory data analysis, where
users expect a relatively short response time.

As an alternative to our relational Bayesian sets approach, consider
the following direct modification of the
standard Bayesian sets formulation to this problem: merge the data sets
$\mathcal D_A$ and $\mathcal D_B$ into a single data set, creating for
each pair $(A^i, B^j)$ a row in the database with an extra binary
indicator of relationship existence. Create a joint model for pairs by
using the marginal models for $\mathcal A$ and~$\mathcal B$ and
treating different rows as being independent. This ignores the fact
that the resulting merged data points are not really i.i.d. under such
a model, because the same object might appear in multiple relations
[\citet{dzeroski01}]. The model also fails to capture the dependency
between $A^i$ and $B^j$ that arises from conditioning on $L^{ij}$,
even if $A^i$ and $B^j$ are marginally independent. Nevertheless,
heuristically this approach can sometimes produce good results, and
for several types of probability families it is very computationally
efficient. We evaluate it in Section~\ref{secexp-webkb}.

\subsection{Choice of features and relational discrimination}

Our setup assumes that the feature space $\Phi$ provides a reasonable
classifier to predict the existence of links. Useful predictive
features can also be generated automatically with a variety of
algorithms [e.g., the ``structural logistic regression''
of \citet{popescul03}]. See also \citet{dzeroski01}. \citet{jensen02}
discuss shortcomings of methods for automated feature
selection in relational classification.

We also assume feature spaces are the same for all possible
combinations of objects. This allows for comparisons between, for example,
cells from different species, or web pages from different web domains,
as long as features are generated by the same function $\Phi(\cdot,
\cdot)$. In general, we would like to relax this requirement, but for
the problem
to be well-defined, features from the different spaces must be related
somehow. A hierarchical Bayesian formulation for linking different
feature spaces is one possibility which might be treated in a future
work.

\subsection{Priors}

The choice of prior is based on the observed data, in a way that is
equivalent to the choice of priors used in the original formulation of
Bayesian sets [\citet{gha05}]. Let $\widehat{\Theta}$ be the maximum
likelihood estimator of $\Theta$ using the relational database
$(\mathcal D_A, \mathcal D_B, \mathcal L_{AB})$. Since the number of
possible pairs grows at a quadratic rate with the number of objects,
we do not use the whole database for maximum likelihood
estimation. Instead, to get $\widehat{\Theta}$, we use all linked pairs
as members of the ``positive'' class ($L = 1$), and subsample unlinked
pairs as members of the ``negative'' class ($L = 0$). We subsample by
sampling each object uniformly at random from the respective data sets
$\mathcal D_A$ and $\mathcal D_B$ to get a new pair. Since link
matrices $\mathcal L_{AB}$ are usually very sparse, in practice, this
will almost always provide an unlinked pair. Sections
\ref{secexp-webkb} and~\ref{secexp-ppi} provide more details.

We use the prior $P(\Theta) = \mathcal N(\widehat{\Theta},
(c\widehat{\mathbf{T}})^{-1}$), where $\mathcal N(\mathbf{m},
\mathbf{V})$ is a normal of mean~$\mathbf{m}$ and variance
$\mathbf{V}$. Matrix $\widehat{\mathbf{T}}$ is the empirical second
moments matrix of the linked object features, although a different
choice might be adequate for different applications. Constant $c$ is a
smoothing parameter set by the user. In all of our experiments we set
$c$ to be equal to the number of positive pairs. A good choice of $c$
might be important to obtain maximum performance, but we leave this
issue as future work. \citet{wang09} present some sensitivity analysis
results for a
particular application in text analysis.

Empirical priors are a sensible choice, since this is a retrieval, not
a predictive, task. Basically, the entire data set is the
population, from which prior information is obtained on possible query
sets. A data-dependent prior based on the population is
important for an approach such as Bayesian sets, since deviances from
the ``average'' behavior in the data are useful to discriminate
between subpopulations.

\subsection{On continuous and multivariate relations}

Although we focus on measuring similarity of qualitative
relationships, the same idea could be extended to \textit{continuous} (or
ordinal) measures of relationship, or relationships where each
$L^{ij}$ is a vector. For instance, \citet{turney05} measure
relations between words by their co-occurrences on the neighborhood of
specific keywords, such as the frequency of two words being connected
by a specific preposition in a large body of text documents. Several
similarity metrics can be defined on this vector of continuous
relationships. However, given data on word features, one
can easily modify our approach by substituting the logistic regression
component with some multiple regression model.

\section{Ranking hyperlinks on the web}
\label{secexp-webkb}

In the following application we consider a collection of web pages
from several universities: the WebKB collection, where relations are
given by hyperlinks
[\citet{craven98}]. Web pages are classified as being of type
\textit{course}, \textit{department}, \textit{faculty}, \textit{project},
\textit{staff},
\textit{student} or \textit{other}. Documents come from four universities
(\textit{Cornell}, \textit{Texas}, \textit{Washington} and \textit
{Wisconsin}). We
are interested in recovering pairs of web pages $\{A, B\}$ where
web page $A$ has a link to web page $B$. Notice that the relationship is
asymmetric. Different types of web pages imply different types of
links. For instance, a \textit{faculty} web page linking to a \textit{
project} web page constitutes a type of link. The analogical reasoning
task here is simplified if we assume each web page object has a single
role (i.e., exactly one out of the pre-defined types
\{\textit{course}, \textit{department}, \textit{faculty},
\textit{project}, \textit{staff}, \textit{student}, \textit{other}\}),
and therefore a pair of web pages implies a unique type of
relationship. The web page types are for evaluation purposes only,
as we explain later: we will not provide this information to the model.

Our main standard of comparison is a ``flattened Bayesian sets''
algorithm (which we will call ``standard Bayesian sets,'' \textsc
{SBSets}, in constrast to the relational model, \textsc{RBSets}). Using a
multivariate independent Bernoulli model as in the original paper
[\citet{gha05}], we merge linked web page pairs into single rows, and
then apply
the original algorithm directly to the merged data. It is clear that
data points are not independent anymore, but the \textsc{SBSets} algorithm
assumes this is the case. Evaluating this algorithm serves the purpose
of both
measuring the loss of not treating relational data as such, as well as
the limitations of evaluating the similarity of pairs through models for
the marginal probabilities of $\mathcal A$ and $\mathcal B$ instead of
models for the predictive function $P(L^{ij}\vert X^{ij})$.

Binary data was extracted from this database using the same methodology
as in
\citet{gha05}. A total of 19,450 binary variables per object are
generated, where each variable indicates whether a word from a fixed
dictionary appears in a given document more frequently than the
average. To avoid introducing extra approximations into \textsc{RBSets},
we reduced the dimensionality of the original representation using
singular value decomposition, obtaining 25 measures per object.

In this experiment objects are of the same type, and therefore,
dimensionality. The feature vector $X^{ij}$ for each pair of objects
$\{A^i, B^j\}$ consists of the $V$ features for object $A^i$, the $V$
features of object $B^j$, and measures $\mathbf Z = \{Z_1, \ldots,
Z_V\}$, where $Z_v = (A^i_v \times B^j_v) / (\vert A^i\vert \times\|B^j\|)$,
$\|A^i\|$ being the Euclidean norm of the $V$-dimensional
representation of $A^i$. We also add a constant value (1) to the
feature set as an intercept term for the logistic regression. Feature
set $\mathbf{Z}$ is exactly the one used in the cosine distance
measure,\footnote{The cosine similarity measure between two items
corresponds to the sum of the features in $\mathbf{Z}$.} a common and
practical measure widely used in information retrieval
[\citet{manning08}]. This feature space also has the important
advantage of scaling well (linearly) with the number of variables in
the database. Moreover, adopting such features will make our
comparisons fairer, since we evaluate how well cosine distance
itself performs in our task. Notice that our choice of $X^{ij}$
is suitable for asymmetric relationships, as naturally occurs in the domain
of web page links. For symmetric relationships, features such as
$|A^i_v - B^j_v|$ could be used instead.

In order to set the empirical prior, we sample 10 ``negative'' pairs
for each ``positive'' one, and weight them to reflect the proportion
of linked to unlinked pairs in the database. That is, in the WebKB
study we use 10 negatives for each positive, and we count each negative
case as being 350 cases replicated. We perform subsampling and
reweighting in order to be able to fit the database in the memory of a
desktop computer.

Evaluation of the significance of retrieved items often relies on
subjective assessments [\citet{gha05}]. To simplify our study, we will
focus on particular setups where objective measures of success are
defined.

\begin{figure}

\includegraphics{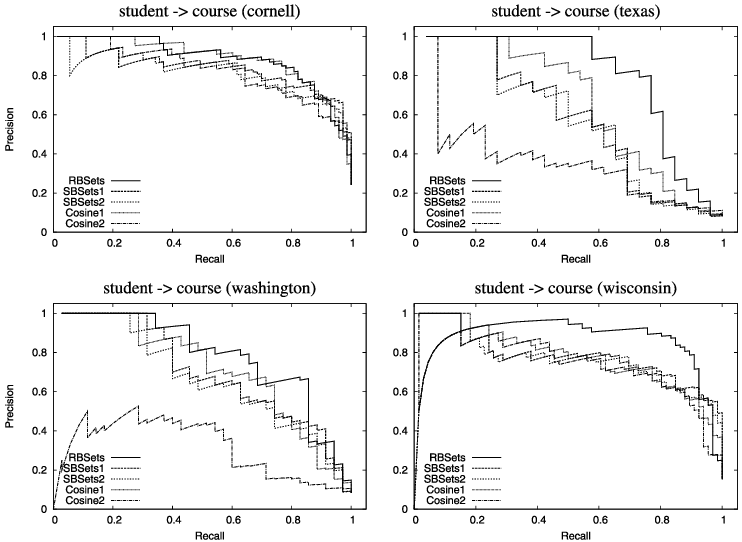}

\caption{Results for \textit{student $\rightarrow$ course} relationships.}
\label{figsc-exp}
\end{figure}

To evaluate the gain of our model over competitors, we will use the
following setup. In the first query, we are given all pairs of
web pages of the type \textit{student $\rightarrow$ course} from three of
the labeled universities, and evaluate how relations are ranked in the
fourth university. Because we know class labels for the web pages
(while the algorithm
does not), we can use the classes of the returned pairs to label a hit
as being ``relevant'' or ``irrelevant.'' We label a pair $(A^i, B^j)$
as relevant if and only if $A^i$ is of type \textit{student} and $B^j$ is
of type \textit{course}, and $A^i$ links to $B^j$.

This is a very stringent criterion, since other types of relations
could also be valid (e.g., \textit{staff $\rightarrow$ course} appears to
be a reasonable match). However, this facilitates objective
comparisons of algorithms. Also, the \textit{other} class contains many
types of pages, which allows for possibilities such as a \textit{student
$\rightarrow$ ``hobby''} pair. Such pairs might be hard to evaluate
(e.g., is that particular hobby demanding or challenging
in a similar way to coursework?). As a compromise, we omit all pages from
the category \textit{other} in order to better clarify differences
between algorithms.\footnote{As an extreme example, querying \textit{
student} $\rightarrow$ \textit{course} pairs from the \textit{wisconsin}
university returned \textit{student} $\rightarrow$ \textit{other} pairs at
the top four. However, these \textit{other} pages were for some reason
course pages---such as \url{http://www.cs.wisc.edu/\textasciitilde markhill/cs752.html}.}

Precision/recall curves [\citet{manning08}] for the \textit{student
$\rightarrow$ course} queries are shown in Figure\vadjust{\goodbreak}
\ref{figsc-exp}. There are four queries, each corresponding to a
search over a specific university
given all valid \textit{student} $\rightarrow$ \textit{course} pairs from
the other three. There are four algorithms on each evaluation: the
standard Bayesian sets with the original 19,450 binary variables for
each object, plus another 19,450 binary variables, each corresponding
to the product of the respective variables in the original pair of
objects (\textsc{SBSets1}); the standard Bayesian sets with the original
binary variables only (\textsc{SBSets2}); a standard cosine distance
measure over the 25-dimensional representation (\textsc{Cosine 1}) for each
page, with pairs being given by the combined vector of 50 features; a
cosine distance measure using the raw 19,450-dimensional binary for
each document (\textsc{Cosine 2}); our approach, \textsc{RBSets}.

\begin{figure}

\includegraphics{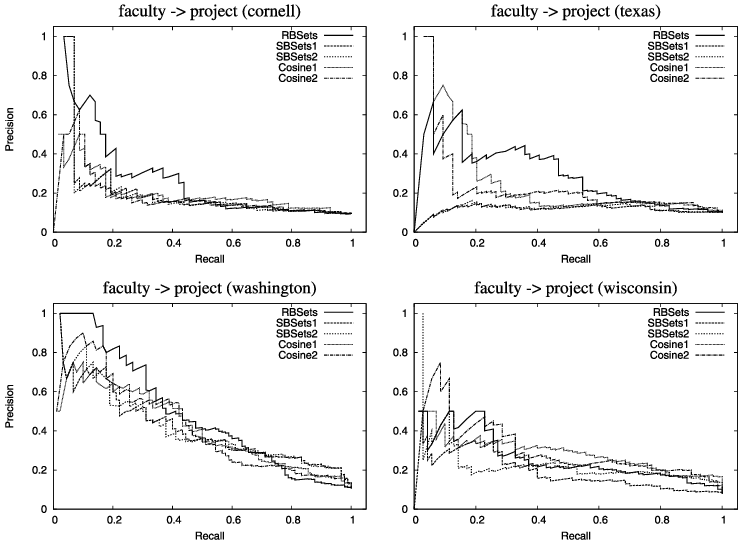}

\caption{Results for \textit{faculty $\rightarrow$ project} relationships.}
\label{figfp-exp}
\end{figure}

In Figure~\ref{figsc-exp} \textsc{RBSets} demonstrates consistently
superior or equal precision-recall. Although \textsc{SBSets} performs
well when asked to retrieve only \textit{student} items or only
\textit{course} items, it falls short of detecting what features of
\textit{
student} and \textit{course} are relevant to predict a link. The
discriminative model within \textsc{RBSets} conveys this
information through the link parameters.

We also did an experiment with a query of type \textit{faculty
$\rightarrow$ project}, shown in Figure~\ref{figfp-exp}. This time
results between algorithms were closer to each other. To make
differences more
evident, we adopt a slightly different measure of success:
we count as a 1 hit if the pair retrieved is a \textit{faculty
$\rightarrow$ project} pair, and count as a 0.5~hit for pairs of type
\textit{student $\rightarrow$ project} and \textit{staff $\rightarrow$
project}. Notice this is a much harder query. For instance, the
structure of the \textit{project} web pages in the \textit{texas} group was
quite distinct from the other universities: they are mostly very
short, basically containing links for members of the project and other
project web pages.

Although the precision/recall curves convey a global picture of
the performance of each algorithm, they might not be a completely clear way
of ranking approaches for cases where curves intersect at several
points. In order to summarize algorithm performances with a single
statistic, we computed the area under each precision/recall curve
(with linear interpolation between points). Results are given in Table
\ref{tabauc}. Numbers in bold indicate the largest area under the curve.
The dominance of \textsc{RBSets} should be clear.

%
\begin{table}
\caption{Area under the precision/recall curve for each algorithm and query}
\label{tabauc}
\begin{tabular*}{\textwidth}{@{\extracolsep{\fill}}lcccccccccc@{}}
\hline
& \multicolumn{5}{c}{\textbf{\textit{Student}} $\bolds{\rightarrow}$ \textbf{\textit{course}}} & \multicolumn{5}{c@{}}{\textbf{\textit{Faculty}}
$\bolds{\rightarrow}$ \textbf{\textit{project}}}\\[-6pt]
& \multicolumn{5}{c}{\hrulefill} & \multicolumn{5}{c@{}}{\hrulefill}\\
&\textbf{C1} & \textbf{C2} & \textbf{RB} & \textbf{SB1} & \textbf{SB2}
& \textbf{C1} &
\textbf{C2} & \textbf{RB} & \textbf{SB1} & \textbf{SB2}\\
\hline
Cornell & \textbf{0.87} & 0.82 & \textbf{0.87} & 0.82 & 0.80 & 0.19 & 0.18 &
\textbf{0.24} & 0.18 & 0.18\\
Texas & 0.62 & 0.32 & \textbf{0.77} & 0.55 & 0.54 & 0.24 & 0.21 & \textbf{
0.29} & 0.12 & 0.12\\
Washington & 0.69 & 0.31 & \textbf{0.76} & 0.67 & 0.64 & 0.40 & 0.42 &
\textbf{0.47} & 0.40 & 0.40\\
Wisconsin & 0.77 & 0.72 & \textbf{0.88} & 0.75 & 0.73 & 0.28 & \textbf{0.30}
& 0.26 & 0.19 & 0.21\\
\hline
\end{tabular*}
\end{table}

\section{Ranking protein interactions}
\label{secexp-ppi}

The budding yeast is a unicellular organism that has become a
de-facto model organism for the study of molecular and cellular
biology [\citet{BotsCherCher1997}]. There are about 6000~proteins
in the budding yeast, which interact in a number of ways
[\citet{CherBallWengJuvi1997}]. For instance, proteins bind
together to form protein complexes, the physical units that carry out
most functions in the cell [\citet{KrogCagnYuZhon2006}]. In recent
years, significant resources have been directed to collect
experimental evidence of physical proteins binding, in an effort to
infer and catalogue protein complexes and their multifaceted
functional roles [e.g., \citet{FielSong1989};
\citet{ItoTashMutaOzaw2000}; \citet{UetzGiotCagnMans2000};
\citet{GaviBoscKrauGran2002}; \citet{HoGruhHeilBade2002}].
Currently,
there are four main sources of interactions between pairs of proteins
that target proteins localized in different cellular compartments with
variable degrees of success: (i) literature curated interactions
[\citet{ReguBreiBoucBrei2006}], (ii) yeast two-hybrid (Y2H)
interaction assays [\citet{YuBrauYildLemm2008}], (iii) protein
fragment complementation (PCA) interaction assays
[\citet{TaraMessLandRadi2008}], and (iv) tandem affinity
purification (TAP) interaction assays
[\citet{GaviAloyGranKrau2006}; \citet{KrogCagnYuZhon2006}]. These
collections include a total of about 12,292 protein interactions
[\citet{JensBork2008}], although the number of such interactions is
estimated to be between 18,000 [\citet{YuBrauYildLemm2008}] and
30,000 [\citet{vonMKrauSnelCorn2002}].

Statistical methods have been developed for analyzing many aspects of
this large protein interaction network, including de-noising
[\citet{BernVaugHart2007}; \citet{AiroBleiFienXing2008}], function
prediction [\citet{NabiJimAgarChaz2005}] and identification of
binding motifs [\citet{BankNabiPeteSing2008}].

\subsection{Overview of the analysis}

We consider multiple functional categorization systems for the
proteins in budding yeast. For evaluation purposes, we use individual
proteins' functional annotations curated by the Munich Institute for
Protein Sequencing [MIPS, \citet{mewes04}], those by the Kyoto
Encyclopedia of Genes and Genomes [KEGG, \citet{kanegoto2000}] and
those by the Gene Ontology consortium [GO, \citet{Ashbetal2000}].
We consider multiple collections of physical protein interactions that
encode alternative semantics. Physical protein-to-protein interactions
in the MIPS curated collection measure physical binding events
observed experimentally in Y2H and TAP experiments, whereas physical
protein-to-protein interactions in the KEGG curated collection measure
a number of different modes of interactions,\vadjust{\goodbreak} including phosporelation,
methylation and physical binding, all taking place in the context of a
specific signaling pathway. So we have three possible functional
annotation databases (MIPS, KEGG and GO) and two possible link
matrices (MIPS and KEGG), which can be combined.

Our experimental pipeline is as follows:
(i) Pick a database of functional annotations, say, MIPS, and a
collection of interactions, say, MIPS (again).
(ii) Pick a pair of categories, $M_1$ and $M_2$. For instance, take
$M_1$ to be \textit{cytoplasm} (MIPS~40.03) and $M_2$ to be \textit{
cytoplasmic and nuclear degradation} (MIPS 06.13.01).
(iii) Sample, uniformly at random and without replacement, a set
$\mathbf{S}$ of 15 interactions in the chosen collection.
(iv) Rank other interacting pairs\footnote{The portion of ranked list
that is relevant for evaluation purposes is limited to a subset of the
protein--protein interactions. More details are given in Section \protect\ref
{secppi-eval}.} according to the score in equation (\ref{eqscore}) and,
for comparison purposes, according to three other approaches to be
described in Section~\ref{secalgo}.
(v) The process is repeated for a large number of pairs $M_1 \times
M_2$, and 5 different query sets $\mathbf{S}$ are generated for each
pair of categories.
(vi) Calculate an evaluation metric for each query and each of the four
scores, and report a comparative summary of the results.

%
%
\begin{table}
\caption{Collection of data sets used to generate protein-specific features}
\label{tabdata-features}
\begin{tabular*}{\textwidth}{@{\extracolsep{\fill}}lll@{}}
\hline
\textbf{No.} & \textbf{Measurements description} & \textbf{Data sources} \\
\hline
1. & Expression microarrays & \citet{GascSpelKaoCarm2000};
\citet{bremstorwhitkrug2005}; \\
& & \citet{primwillwinztevz2000}; \citet{yverbremwhitakey2003} \\
2. & Synthetic genetic interactions
& \citet{breistartyer2003}; \citeauthor{sgd} \\
3. & Cellular localization & \citet{huhfalvgerkcarr2003} \\
4. & Transcription factor binding sites
& \citet{harbgordleerina2004}; \citeauthor{transfac} \\
5. & Sequence similarities & \citet{altsgishmille1990}; \citet{zhuzhan1999} \\
\hline
\end{tabular*}
\end{table}
%

\subsubsection{Protein-specific features}

The protein-specific features were generated using the data sets
summarized in Table~\ref{tabdata-features} and an additional data set
[\citet{qi06}]. Twenty gene expression attributes were obtained from
the data set processed by \citet{qi06}. Each gene expression attribute
for a protein pair $P_i\dvtx P_j$ corresponds to the correlation coefficient
between the expression levels of corresponding genes. The 20 different
attributes are obtained from 20 different experimental conditions as
measured by microarrays. We did not use pairs of proteins from Qi et
al. for which we did not have data in the data sets listed in Table
\ref{tabdata-features}. This resulted in approximately 6000
positively linked data points for the MIPS network and 39,000 for
KEGG.

We generated another 25 protein--protein gene expression features from
the data in Table~\ref{tabdata-features} using the same procedure
based on correlation coefficients. This gives a total of 45
attributes, corresponding to the main data set used in our relational
Bayesian sets runs.

Another data set was generated using the remaining (i.e.,
nonmicroarray) features of Table~\ref{tabdata-features}. Such
features are binary and highly sparse, with most entries being 0 for
the majority of linked pairs. We removed attributes for which we
had fewer than 20 linked pairs with positive values according to the
MIPS network. The total number of extra binary attributes was 16.

Several measurements were missing. We imputed missing values for each
variable in a particular data point by using its empirical average
among the observed values.

Given the 45 or 61 attributes of a given pair \{$P_i$, $P_j$\}, we
applied a nonlinear transformation where we normalize the vector by
its Euclidean norm in order to obtain our feature table $\mathbf X$.

\subsubsection{\texorpdfstring{Calibrating the prior for $\Theta$}
{Calibrating the prior for Theta}}

We initially fit a logistic regression classifier using a maximum\vspace*{1pt}
likelihood estimation (MLE) and our data, obtaining the estimate
$\widehat{\Theta}$.
Our choice of covariance matrix $\widehat{\Sigma}$ for $\Theta$ is
defined to be a rescaling of
a squared norm of the data:
%
%
\begin{equation}
(\widehat{\Sigma})^{-1} = \mathbf{X}^\mathsf{T}_{\mathrm{POS}}\mathbf{X}_{\mathrm{POS}},
\label{eqcov}
\end{equation}

\noindent where $\mathbf{X}_{\mathrm{POS}}$ is the matrix containing the protein--protein
features only of the linked pairs used in the MLE computation.

\subsubsection{Evaluation metrics}
\label{secppi-eval}

As in the WebKB experiment, we propose an objective measure of
evaluation that is used to compare different algorithms.
Consider a query set $\mathbf{S}$, and a ranked response list $\mathbf
{R} =
\{R^1, R^2, R^3, \ldots,R^N\}$ of protein--protein pairs. Every
element of
$\mathbf{S}$ is a pair of proteins $P_i\dvtx P_j$ such that~$P_i$ is of
class $M_i$ and $P_j$ is of class $M_j$, where $M_i$ and $M_j$ are
classes from either MIPS, KEGG or Gene Ontology. In general, proteins
belong to multiple classes. This is in contrast with the WebKB
experiment, where, according to our web page categorization, there was
only one possible type of relationship for each pair of web pages.
The retrieval algorithm that generates $\mathbf{R}$ does not receive
any information concerning the MIPS, KEGG or GO taxonomy. $\mathbf{R}$
starts with the linked protein pair that is judged most similar to
$\mathbf{S}$, followed by the other protein pairs in the population,
in decreasing order of similarity. Each algorithm has its own measure
of similarity.

The evaluation criterion for each algorithm is as follows: as before,
we generate a precision-recall curve and calculate the area under the
curve (AUC). We also calculate the proportion (TOP10), among the top
10 elements in each ranking, of pairs that match the original $\{M_1,
M_2\}$ selection (i.e., a ``correct'' $P_i\dvtx P_j$ is one where
$P_i$
is of class $M_1$ and $P_j$ of class $M_2$, or vice-versa.\vadjust{\goodbreak} Notice that
each protein belongs to multiple classes, so both
conditions might be satisfied.) Since a researcher is only likely to
look at the top ranked pairs, it makes sense to define a measure that
uses only a subset of the ranking. AUC and TOP10 are our two
evaluation measures.

The original classes $\{M_1, M_2\}$ are known to the experimenter but
not known to the algorithms. As in the WebKB experiment, our criterion
is rather stringent, in the sense that it requires a perfect match of each
$R^I$ with the MIPS, KEGG or GO categorization.
There are several ways by which a pair $R^I$ might be analogous to the
relation implicit in $\mathbf{S}$, and they do not need to agree with
MIPS, GO or KEGG. Still, if we are willing to believe that these
standard categorization systems capture functional organization of
proteins at some level, this must lead to association between
categories given to $\mathbf{S}$ and relevant subpopulations of
protein--protein interactions similar to $\mathbf{S}$. Therefore, the
corresponding AUC and TOP10 are useful tools for comparing
different algorithms even if the actual measures are likely to be
pessimistic for a fixed algorithm.

\subsubsection{Competing algorithms}
\label{secalgo}

We compare our method against a variant of it and two similarity
metrics widely used for information retrieval:%
\begin{enumerate}
\item The cosine score [\citet{manning08}], denoted by \textsc{cos}.
\item The nearest neighbor score, denoted by \textsc{nns}.
\item The relational maximum likelihood sets score, denoted by \textsc{mls}.
\end{enumerate}
The nearest neighbor score measures the minimum Euclidean distance
between~$R^I$ and any individual point in $\mathbf{S}$, for a given
query set $\mathbf{S}$ and a given candidate point~$R^I$.
The relational maximum likelihood sets is a variation of \textsc{RBSets}
where we initially sample a subset of the
unlinked pairs (10,000 points in our setup) and, for each query
$\mathbf{S}$, we fit a logistic regression model to obtain the
parameter estimate $\Theta_{\mathbf{S}}^{\mathrm{MLE}}$. We also use a logistic
regression model fit to the \textit{whole} data set (the same one used to
generate the prior for \textsc{RBSets}), giving the estimate
$\Theta^{\mathrm{MLE}}$. A new score, analogous to (\ref{eqscore}), is given
by $\log P(L^{ij} = 1 | X^{ij}, \Theta_{\mathbf{S}}^{\mathrm{MLE}}) - \log
P(L^{ij} = 1 | X^{ij}, \Theta^{\mathrm{MLE}})$, that is, we do not integrate out
the parameters or use a prior, but instead the models are fixed at
their respective estimates.

Neither \textsc{cos} or \textsc{nns} can be interpreted as measures of
analogical similarity, in the sense that they do not take into account how
the protein pair features $\mathbf X$ contribute to their
interaction.\footnote{As a consequence, none uses negative
data. Another consequence is the necessity of modeling the input space
that generates $\mathbf X$, a difficult task given the dimensionality
and the continuous nature of the features.} It is true that a direct
measure of analogical similarity is not theoretically required
to
perform well according to our (nonanalogical) evaluation
metric. However, we will see that there are practical advantages in doing
so.\vadjust{\goodbreak}

\subsection{Results on the MIPS collection of physical interactions}
\label{seceval-mips}

For this batch of experiments, we use the MIPS network of
protein--protein interactions to define the relationships. In the
initial experiment, we selected queries from all combinations of MIPS
classes for which there were at least 50 linked pairs $P_i\dvtx P_j$ in the
network that satisfied the choice of classes. Each query set contained
15 pairs.
After removing the MIPS-categorized proteins for which we had no
feature data,
we ended up with a total of 6125 proteins
and 7788 positive interactions. We set the prior for \textsc{RBSets}
using a sample of 225,842 pairs labeled as having no interaction,
as selected by \citet{qi06}.

For each tentative query set $\mathbf{S}$ of categories $\{M_1, M_2\}$,
we scored and ranked pairs $P_i'\dvtx P_j'$ such that both $P_i'$ and
$P_j'$ were connected to some protein appearing in~$\mathbf{S}$ by a
path of no more than two steps, according to the MIPS network. The
reasons for the filtering are two-fold: to increase the
computational performance of the ranking since fewer pairs are scored;
and to minimize the chance that undesirable pairs would appear in the
top 10 ranked pairs. Tentative queries would not be performed if after
filtering we obtained fewer than 50 possible correct matches. Trivial
queries, where filtering resulted only in pairs in the same class as
the query, were also discarded.
The resulting number of unique pairs of categories $\{M_1, M_2\}$ was
931 classes of interactions. For each pair of categories, we sampled
our query set $\mathbf{S}$ 5 times, generating a total of 4655 rankings
per algorithm.
%
%

\begin{table}
\caption{Number of times each method wins when querying pairs of
MIPS classes using the MIPS protein--protein interaction network. The
first two columns, \#\textsc{AUC} and \#\textsc{TOP10}, count the number of
times the respective method obtains the best score
according to the AUC and TOP10 measures, respectively, among the 4
approaches. This is divided by the number of replications of each
query type (5). The last two columns, \#\textsc{AUC.S} and \#\textsc{TOP10.S}, are
``smoothed'' versions of this statistic: a method is
declared the winner of a round of 5 replications if it obtains the
best score in at least 3 out of the 5 replications. The top table
shows the results when only the continuous variables are used by \textsc{RBSets},
and in the bottom table when the discrete variables are also given
to \textsc{RBSets}}
\label{tabmips-mips}
\begin{tabular*}{\textwidth}{@{\extracolsep{\fill}}lcccc@{}}
\hline
\textbf{Method} & \textbf{\#AUC} & \textbf{\#TOP10} & \textbf{\#AUC.S} & \textbf{\#TOP10.S}
\\
\hline
\multicolumn{5}{c}{(a)}\\
{COS} & 240 & 294 & 219 & 277 \\
{NNS} & \phantom{0}42 & 122 & \phantom{0}28 & \phantom{0}75 \\
{MLS} & 105 & 270 & \phantom{0}52 & 198 \\
\textsc{RBSets} & 542 & 556 & 578 & 587 \\[6pt]
\multicolumn{5}{c}{(b)}\\
{COS} & 314 & 356 & 306 & 340 \\
{NNS} & \phantom{0}75 & 146 & \phantom{0}62 & 111 \\
{MLS} & 273 & 329 & 246 & 272 \\
\textsc{RBSets} & 267 & 402 & 245 & 387 \\
\hline
\end{tabular*}
\end{table}

%
\begin{table}[b]
\caption{Pairwise comparison of methods according to the AUC and TOP10
criterion.\break Each cell shows the proportion of the trials where the
method in the respective row wins~over~the method in the column,
according to both criteria. In each cell, the proportion is
calculated with respect to the 4655 rankings where no tie happened}
\label{tabmips-pairwise}
\begin{tabular*}{\textwidth}{@{\extracolsep{\fill}}lcccccccc@{}}
\hline
& \multicolumn{4}{c}{\textbf{AUC}} & \multicolumn{4}{c@{}}{\textbf{TOP10}}\\[-6pt]
& \multicolumn{4}{c}{\hrulefill} & \multicolumn{4}{c@{}}{\hrulefill}\\
& \textbf{COS} & \textbf{NNS} & \textbf{MLS} & \textbf{\textsc{RBSets}} &
\textbf{COS} & \textbf{NNS} & \textbf{MLS} & \textbf{\textsc{RBSets}}\\
\hline
\textsc{COS} & -- & 0.67 & 0.43 & 0.30 & -- & 0.70 & 0.46 & 0.30 \\
\textsc{NNS} & 0.32 & -- & 0.18 & 0.06 & 0.29 & -- & 0.25 & 0.11 \\
\textsc{MLS} & 0.56 & 0.81 & -- & 0.25 & 0.53 & 0.74 & -- & 0.28 \\
\textsc{RBSets} & 0.69 & 0.93 & 0.74 & -- & 0.69 & 0.88 & 0.71 & -- \\
\hline
\end{tabular*}
\end{table}

We run two types of experiments. In one version, we give to \textsc{RBSets}
the data containing only the 45 (continuous) microarray measurements.
In the
second variation, we provide to \textsc{RBSets} all 61 variables,
including the
16 sparse binary indicators. However, we noticed that the addition of
the 16
binary variables hurts \textsc{RBSets} considerably. We conjecture that
one reason
might be the degradation of the variational approximation. Including
the binary
variables hardly changed the other three methods, so we choose to use the
61 variable data set for the other methods.\footnote{We also performed
an experiment (not
included) where only the continuous attributes were used by the other
methods. The advantage of \textsc{RBSets} still increased, slightly (by a
2\% margin against the cosine distance method). For this reason, we
analyze the most pessimistic case.}

Table~\ref{tabmips-mips} summarizes the results of this
experiment. We show the number of times each method wins according to
both the AUC and TOP10 criteria. The number of wins is presented as
divided by 5, the number of random sets generated for each query type
$\{M_1, M_2\}$ (notice these numbers do not need to add up to 931,
since ties are possible). Moreover, we also presented ``smoothed''
versions of this statistic, where we count a method as the winner for
any given $\{M_1, M_2\}$ category if, for the group of 5 queries, the
method obtains the best result in at least 3 of the sets. The
motivation is to smooth out the extra variability added by the
particular set of 15 protein pairs for a fixed $\{M_1, M_2\}$. The
proposed relational Bayesian sets method is the clear winner according
to all measures when we select only the continuous variables. For this
reason, for the rest of this section all analysis and experiments will
consider only this case.

Table~\ref{tabmips-pairwise} displays a pairwise comparison of the
methods. In this table we show how often the row method performs better than
the column method, among those trials where
there was no tie. Again, \textsc{RBSets} dominates.

Another useful summary is the distribution of correct hits in the top 10
ranked elements across queries. This provides a measure of the difficulty
of the problem, besides the relative performance of each algorithm.
In Table~\ref{tabmips-top10} we show the proportion of
correct hits among the top 10 for each algorithm for our queries using
MIPS categorization and also GO categorization, as explained in the
next section. About 14\% of the time, all pairs in the top 10 pairs
ranked by \textsc{RBSets} were of the intended type, compared to 8\% of the
second best approach.


%
\begin{table}
\def\arraystretch{0.9}
\caption{Distribution across all queries of the number of hits in the
top 10 pairs,\break
as ranked by each algorithm. The more skewed to the right, the better.
Notice that
using GO categories doubles the number of zero hits for \textsc{RBSets}}\label{tabmips-top10}
\begin{tabular*}{\textwidth}{@{\extracolsep{\fill}}lccccccccccc@{}}
\hline
& \textbf{0} & \textbf{1} & \textbf{2} & \textbf{3} & \textbf{4}
&\textbf{5} & \textbf{6} & \textbf{7} & \textbf{8} & \textbf{9} & \textbf{10}\\
\hline
\multicolumn{12}{c}{Proportion of top hits using MIPS categories and links specified by the
MIPS database}\\
\textsc{COS} & 0.12 &0.15 &0.12 &0.10 &0.08 &0.07 &0.06 &0.05 &0.04 &0.07
&0.08 \\
\textsc{NNS} & 0.29 &0.16 &0.14 &0.10 &0.06 &0.05 &0.03 &0.03 &0.03 &0.03
&0.02 \\
\textsc{MLS} & 0.12 &0.12 &0.12 &0.10 &0.09 &0.08 &0.07 &0.06 &0.07 &0.06
&0.07 \\
\textsc{RBSets} & 0.04 &0.08 &0.09 &0.09 &0.09 &0.08 &0.09 &0.07 &0.09
&0.08 &0.14 \\[6pt]
\multicolumn{12}{c}{Proportion of top hits using GO categories and links specified by the
MIPS database}\\
\textsc{COS} & 0.12 &0.13 &0.11 &0.10 &0.11 &0.09 &0.06 &0.06 &0.04 &0.06
&0.06 \\
\textsc{NNS} & 0.53 &0.23 &0.07 &0.02 &0.02 &0.02 &0.04 &0.01 &0.00 &0.00
&0.01 \\
\textsc{MLS} & 0.16 &0.11 &0.12 &0.10 &0.08 &0.08 &0.08 &0.06 &0.05 &0.06
&0.05 \\
\textsc{RBSets} & 0.09 &0.09 &0.10 &0.10 &0.08 &0.08 &0.06 &0.08 &0.08
&0.07 &0.12 \\
\hline
\end{tabular*}
\end{table}

%
\begin{table}[b]
\def\arraystretch{0.9}
\caption{Number of times each method wins when querying pairs of
GO classes using the MIPS protein--protein interaction network. Columns
\#\textsc{AUC},
\#\textsc{TOP10}, \#\textsc{AUC.S} and \#\textsc{TOP10.S} are defined as in Table
\protect\ref{tabmips-mips}}
\label{tabgo-mips}
\begin{tabular*}{\textwidth}{@{\extracolsep{\fill}}lcccc@{}}
\hline
\textbf{Method} & \textbf{\#AUC} & \textbf{\#TOP10} & \textbf{\#AUC.S} & \textbf{\#TOP10.S}
\\
\hline
\textsc{COS} & 58 & \phantom{0}73 & \phantom{0}58 & \phantom{0}72 \\
\textsc{NNS} & \phantom{0}1 & \phantom{0}10 & \phantom{00}0 & \phantom{00}4 \\
\textsc{MLS} & 26 & \phantom{0}55 & \phantom{0}13 & \phantom{0}38 \\
\textsc{RBSets} & 93 & \phantom{.}105 & \phantom{.}101 & 110 \\
\hline
\end{tabular*}
\end{table}

\subsubsection{Changing the categorization system}

A variation of this experiment was performed where the protein
categorizations do \textit{not} come from the same family as the link
network, that is, where we used the MIPS network but not the MIPS
categorization. Instead we performed queries according to the Gene
Ontology categories. Starting from 150 pre-selected GO categories
[\citet{MyerBarrHibbHutt2006}], we once again generated unordered
category pairs $\{M_1, M_2\}$. A total of 179 queries, with
5 replications each (a total of 895 rankings), were generated and
the results summarized in Table~\ref{tabgo-mips}.\vadjust{\goodbreak}

This is a more challenging scenario for our approach, which is
optimized with respect to MIPS. Still, we are able to outperform other
approaches. Differences are less dramatic, but consistent. In the pairwise
comparison of \textsc{RBSets} against the second best method, \textsc{cos}, our
method wins 62\% of the time by the TOP10 criterion.

\subsubsection{The role of filtering}

In both experiments with the MIPS network, we filtered candidates
by examining only a subset of the proteins linked to the elements in
the query set by a path of no more than two proteins. It is relevant
to evaluate how much coverage of each category pair $\{M_1, M_2\}$ we
obtain by this neighborhood selection.

\begin{figure}

\includegraphics{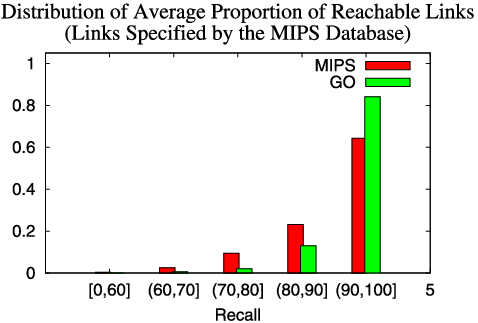}

\caption{Distribution of the coverage of valid pairs in the MIPS network,
according to our generated query sets. Results are broken into the
two categorization systems (MIPS and GO) used in this experiment.}
\label{figmips-dist}
\end{figure}

For each query $\mathbf{S}$, we calculate the proportion of pairs
$P_i\dvtx P_j$ of the same categorization $\{M_1, M_2\}$ such that both
$P_i$ and $P_j$ are included in the neighborhood. Figure
\ref{figmips-dist} shows the resulting distributions of such
proportions (from 0 to 100\%): a histogram for the MIPS search and a
histogram for the GO search. Despite the small neighborhood, coverage
is large. For the MIPS categorization, 93\% of the queries resulted in
a coverage of at least 75\% (with 24\% of the queries resulting in
perfect coverage). Although filtering implies that some valid pairs
will never be ranked, the gain obtained by reducing false positives in
the top 10 ranked pairs is considerable (results not shown) across all
methods, and the computational gain of reducing the search space is
particularly relevant in exploratory data analysis.

\subsection{Results on the KEGG collection of signaling pathways}
\label{seceval-kegg}

We repeat the same experimental setup, now using the KEGG network to
define the protein--protein interactions.\vadjust{\goodbreak} We selected proteins from the
KEGG categorization system for which we had data available. A total
of 6125 proteins were selected. The KEGG network is much more dense
than MIPS. A total of 38,961 positive pairs and 226,188 negative links
were used to generate our empirical prior.

However, since the KEGG network is much more dense than MIPS, we filtered
our candidate pairs by allowing only proteins that are directly linked to
the proteins in the query set $\mathbf{S}$. Even under this restriction, we
are able to obtain high coverage: the neighborhood of 90\% of the queries
included all valid pairs of the same category, and essentially all queries
included at least 75\% of the pairs falling in the same category as the
query set. A total of 1523 possible pairs of categories (7615 queries,
considering the 5 replications) were generated.
%
\begin{table}
\caption{Number of times each method wins when querying pairs of
KEGG classes using the KEGG protein--protein interaction
network. Columns \#\textsc{AUC}, \#\textsc{TOP10}, {\#\textsc{AUC.S} and
\#\textsc{TOP10.S} are defined as in Table
\protect\ref{tabmips-mips}}}\label{tabkegg-kegg}
\begin{tabular*}{\textwidth}{@{\extracolsep{\fill}}lcccc@{}}
\hline
\textbf{Method} & \textbf{\#AUC} & \textbf{\#TOP10} & \textbf{\#AUC.S} & \textbf{\#TOP10.S}
\\
\hline
\textsc{COS} & \phantom{0}159 & \phantom{0}575 & \phantom{0}134 & \phantom{0}507 \\
\textsc{NNS} & \phantom{00}30 & \phantom{0}305 & \phantom{00}17 & \phantom{0}227 \\
\textsc{MLS} & \phantom{0}290 & \phantom{0}506 & \phantom{0}199 & \phantom{0}431 \\
\textsc{RBSets} & 1042 & 1091 & 1107 & 1212 \\
\hline
\end{tabular*}
\end{table}

%
\begin{table}[b]
\caption{Distribution across all queries of the number of hits in the
top 10 pairs,
as ranked by each algorithm. The more skewed to the right, the better}
\label{tabkegg-top10}
\begin{tabular*}{\textwidth}{@{\extracolsep{\fill}}lccccccccccc@{}}
\hline
& \textbf{0} & \textbf{1} & \textbf{2} & \textbf{3} & \textbf{4} & \textbf{5} & \textbf{6} & \textbf{7} & \textbf{8} & \textbf{9} & \textbf{10}\\
\hline
\multicolumn{12}{c}{Proportion of top hits using KEGG categories and links specified by the
KEGG database}\\
\textsc{COS} & 0.56 &0.21 &0.08 &0.03 &0.02 &0.01 &0.01 &0.01 &0.01 &0.01
&0.01 \\
\textsc{NNS} & 0.89 &0.03 &0.04 &0.01 &0.00 &0.00 &0.00 &0.00 &0.00 &0.00
&0.00 \\
\textsc{MLS} & 0.57 &0.21 &0.08 &0.04 &0.02 &0.01 &0.01 &0.00 &0.00 &0.00
&0.00 \\
\textsc{RBSets} & 0.29 &0.24 &0.16 &0.09 &0.06 &0.03 &0.02 &0.01 &0.03
&0.02 &0.01 \\
\hline
\end{tabular*}
\end{table}

Results are summarized in Table~\ref{tabkegg-kegg}. Again, it is
evident that \textsc{RBSets} dominates other methods. In the pairwise
comparison against \textsc{cos}, \textsc{RBSets} wins 76\% of the times
according to the TOP10 criterion. However, the ranking problem in the
KEGG network was much harder than in the MIPS network (according to
our automated nonanalogical criterion). We believe that the reason is
that, in KEGG, the simple filtering scheme has much less influence as
reflected by the high coverage. The distribution of the number of hits
in the top 10 ranked items is shown in
Table~\ref{tabkegg-top10}.
Despite the success of \textsc{RBSets} relative to the
other algorithms, there is room for improvement.


\section{More related work}
\label{secpreviouswork}

There is a large literature on analogical reasoning in artificial
intelligence and psychology. We refer to \citet{french02} for
a survey, and to more recent papers on clustering
[\citet{marx02}], prediction [\citet{turney05}; \citet{turney08}] and dimensionality
reduction [\citet{memisevic05}] as examples of other
applications. Classical approaches for planning have also exploited
analogical similarities [\citet{veloso93}].

Nonprobabilistic similarity functions between relational structures
have also been developed for the purpose of deriving kernel matrices,
such as those required by support vector machines. \citet{karsten07}
provides a comprehensive survey and state-of-the-art methods. It
would be interesting to adapt such methods to problems of analogical reasoning.

The graphical model formulation of \citet{getoor02} incorporates
models of link existence in relational databases, an idea used
explicitly in Section~\ref{secmodel} as the first step of our problem
formulation. In the clustering literature, the probabilistic approach
of \citet{kemp06} is motivated by principles similar to those in our
formulation: the idea is that there is an infinite mixture of
subpopulations that generates the observed relations. Our problem,
however, is to retrieve other elements of a subpopulation described by
elements of a query set, a goal that is closer to the classical
paradigm of analogical reasoning.

As discussed in Section~\ref{secblock-comp},
our model can be interpreted as a type of block model
[\citet{kemp06}; \citet{XuTresYuKrie2006}; \citet{AiroBleiFienXing2008}] with
observable features. Link indicators are independent given the object
features, which might not actually be the case for particular choices
of feature space. In theory, block models sidestep this issue by
learning all the necessary latent features that account for link
dependence. An important future extension of our work would consist of
tractably modeling the residual link association that is not
accounted for by our observed features.


Discovering analogies is a specific task within the general problem of
generating latent relationships from relational data. Some of the
first formal methods for discovering latent relationships from
multiple data sets were introduced in the literature of inductive
logic programming, such as the inverse resolution method
[\citet{muggleton91}]. A more recent probabilistic method is discussed
by \citet{kok07}. \citet{dzeroski01} and \citet{getoor07} provide an
overview of relational learning methods from a data mining and machine
learning perspective.

A particularly active subfield on latent relationship generation lies
within text analysis research. For instance, \citet{stephens01}
describe an approach for discovering relations between genes given
MEDLINE abstracts. In the context of information retrieval,
\citet{cafarella06} describe an application of recent unsupervised
information extraction methods: relations generated from unstructured
text documents are used as a preprocessing step to build an index of
web pages. In analogical reasoning applications, our method
has been used by others for question-answering analysis [\citet{wang09}].

The idea of measuring the similarity of two data points based on a
predictive function has appeared in the literature on matching for
causal inference. Suppose we are given a model for predicting an outcome
$Y$ given a treatment $Z$ and a set of potential confounders $\mathbf
X$. For simplicity, assume $Z \in\{0, 1\}$. The goal of matching is
to find, for each data point $(\mathbf X_i, Y_i, Z_i)$, the closest
match $(\mathbf X_j, Y_j, Z_j)$ according to the confounding variables
$\mathbf X$. In principle, any clustering criterion could be used in
this task [\citet{gelman07}]. The propensity score criterion
[\mbox{\citet{rose02}}] measures the similarity of two feature
vectors~$\mathbf X_i$ and~$\mathbf X_j$ by comparing the
predictions $P(Z_i = 1\vert \mathbf X_i)$ and $P(Z_j = 1\vert \mathbf
X_j)$. If the conditional $P(Z = 1\vert \mathbf X)$ is given by a
logistic regression model with parameter vector $\Theta$,
\citet{gelman07} suggest measuring the difference between $\mathbf
X_i^{\mathsf T}\Theta$
and $\mathbf X_j^{\mathsf T}\Theta$. While this is not the same as
comparing two predictive functions as in our framework, the core idea
of using predictive functions to define similarity remains.

A preliminary version of this paper appeared in the proceedings of the
11th International Conference on Artificial Intelligence and Statistics
[\citet{silva07a}].

\section{Conclusion}

We have presented a framework for performing analogical reasoning
within a Bayesian data analysis formulation. There is of course much
more to analogical reasoning than calculating the similarity of
related pairs. As future work, we will consider hierarchical models
that could in principle compare relational structures (such as protein
complexes) of different sizes. In particular, the literature on graph
kernels [\citet{karsten07}] could provide insights on developing efficient
similarity metrics within our probabilitistic framework.

Also, we would like to combine the properties of the mixed-membership
stochastic block model of \citet{AiroBleiFienXing2008}, where
objects are clustered into multiple roles according to the
relationship matrix $\mathcal L_{AB}$, with our framework where
relationship indicators are conditionally independent given observed
features.

Finally, we would like to consider the case where multiple
relationship matrices are available, allowing for the comparison of
relational structures with multiple types of objects.

Much remains to be done to create a complete analogical reasoning
system, but the described approach has immediate applications to
information retrieval and exploratory data analysis.

\section*{Acknowledgments}
We would like to thank the anonymous reviewers and the editor for several
suggestions that improved the presentation of this paper, and for
additional relevant references.

\begin{supplement}
\sname{Supplement}\label{suppA}
\stitle{Java implementation of the Relational Bayesian Sets method}
\slink[doi]{10.1214/09-AOAS321SUPP}
\slink[url]{http://lib.stat.cmu.edu/aoas/321/supplement.zip}
\sdatatype{.zip}
\sdescription{We provide complete source code for our method, and instructions
on how to rebuild our experiments. With the code it is also possible to
test variations of our queries, analyzing the sensitivity of the results
to different query sizes and initialization of the variational optimizer.}
\end{supplement}

\printaddresses

\end{document}